\documentclass[preprint,prb,showpacs]{revtex4}

\usepackage{graphicx}

\newcommand{\ymo}{YMnO$_3$~}
\newcommand{\gdo}{GdFeO$_3$}

\newcommand{\rmo}{RMnO$_3$~}
\newcommand{\lmo}{LaMnO$_3$~}

\newcommand{\tmo}{TbMnO$_3$~}
\newcommand{\hmo}{HoMnO$_3$~}

\begin{document}




%



\title{First--principles stabilization of an unconventional collinear magnetic 
ordering in distorted manganites}

\author{S. Picozzi}

\affiliation{CNR-INFM, CASTI Regional Lab, I--67010 Coppito (L'Aquila),
Italy}

\author{K. Yamauchi}

\affiliation{ISIR-SANKEN, Osaka University, Mihogaoka 8-1, Ibaraki, Osaka 567-0047, Japan}

\author{G. Bihlmayer and S. Bl\"ugel}

\affiliation{Institut f\"ur Festk\"orperforschung,
Forschungszentrum J\"ulich, 52425 J\"ulich, Germany}
\date{\today}

\begin{abstract}
First-principles calculations have been performed for different
collinear magnetic orderings in orthorhombic manganites, such as \hmo,
\tmo and \ymo, showing large GdFeO$_3$-like distortions.
Our results suggest that the
AFM-E type ordering,  experimentally observed in HoMnO$_3$ and recently proposed from model hamiltonian studies
as a potentially novel phase,
is indeed the magnetic ground state. Its stability is strongly connected with
octahedral distortions and points to the relevance of {\em structural} more than
{\em chemical} effects. The calculated
exchange constants, extracted from a Heisenberg model used to
fit the first-principles
 total energies, show that the ferromagnetic in-plane
 nearest-neighbour coupling is 
reduced compared to less-distorted
manganites, such as \lmo. In parallel, the antiferromagnetic
next-nearest-neighbour coupling along planar Mn-O-O-Mn paths in 
highly-distorted 
manganites plays a relevant role in the stabilization
of the AFM-E spin configuration.
In agreement with experiments,
the density of states shows that this phase is
insulating with an indirect band-gap of $\sim$ 0.5 eV. 
\end{abstract}
\maketitle



\section{Introduction}
Rare-earth manganites show a  
  fascinating variety of physical phenomena,\cite{tokura,rmp} ranging from their unusual magneto-transport properties
 ( leading to the well known
  colossal
magnetoresistance\cite{rmp}) to peculiar charge, spin and orbital orderings to
recently discovered ferroelectricity\cite{kimura,kenzelmann}. It is widely accepted that the intriguing manganites physics
is intimately
related to the profound interplay between the crystal lattice and the  spin,
charge and orbital degrees of freedom.  
In RMnO$_3$ with light rare-earth (R) cations (R= La, Pr, Nd)  the
collinear antiferromagnetic A-type (AFM-A) ordering ({\em i.e.}
ferromagnetically (FM) coupled $ac$ layers antiferromagnetically
 aligned along the $b$ direction in the $Pnma$ setting) is
favored, whereas manganites
toward the end of the series (R= Tb, Dy, Y) exhibit a
sinusoidal magnetic structure defined by the propagation vector ($k_x$,0,0).
Recently, a new AFM phase has been observed  in late-R HoMnO$_3$\cite{munozho} 
as well as in nickelates\cite{nickelates}, with an
``up-up-down-down" spin ordering in the MnO$_2$ planes (the
so called ``E-type" in the
Wollan-Koehler notation\cite{wk}), {\em i.e.} two Mn with up spins are
alternated
by two Mn with down spins along the principal axes of the in--plane cubic
unit--cell, whereas the spins are reversed from plane to plane in the
perpendicular direction. 
It is surprising that
only very
recently this phase has been proposed in undoped manganites\cite{dagotto,kimura1},
considered well--understood systems, where
the E-type ordering was found to be stabilized in a wide region of the phase space.  

The focus of the present work is therefore to 
compare our results with experimental observations suggesting  AFM-E  as the magnetic ground-state in \hmo
and, more in general, to
deeply investigate this non-conventional magnetic ordering, focusing in particular on the link
between magnetic and structural properties. 
In closer detail, we    consider  different
collinear magnetic alignments for three highly distorted manganites, \hmo,
\tmo and \ymo, in  order  to  find  the
stable magnetic configuration and compare our results for the most studied 
\lmo. Surprisingly, we find that the AFM-E is the
ground--state collinear
phase for \hmo, \tmo and \ymo  and that its stabilization is mainly related
to octahedral distortions that decrease the in--plane
FM interaction between $e_g$ states.

\section{Structural and computational details}

The simulations have been performed within the generalized gradient
approximation (GGA) to the exchange--correlation potential in the density-functional
framework. The  all--electron full-potential linearized
augmented plane--wave (FLAPW)\cite{flapw}
 formalism in the FLEUR\cite{fleur} implementation
has been used. Muffin-tin radii have been set to 2.5, 2.5, 2.5, 2.5, 2.0 and 1.5 a.u.
for Tb, Y, La, Ho,
Mn and O atoms, respectively, whereas the wave function cut-off
was chosen as 3.8 a.u.$^{-1}$. Convergency
was carried out using 24 special
{\bf k}-points in the  orthorhombic Brillouin zone,
whereas finer quantities (such as differences
in the total energy) were checked using 192 special
{\bf k}-points.


In order to accurately treat   within  DFT the 4$f$ electrons 
 - that typically occupy   an
unphysical position slightly  above  the  Fermi  level ($E_F$) when considered
as  valence  states -
the Ho and Tb 4$f$ electrons were kept in the core as an {\em 
open shell}
\cite{4f}
(with 7 and 1 (2) electrons in the up and down spin channel, respectively, for Tb (Ho)). An
alternative solution to remove spurious effects related to the wrong energy
position of Tb 4$f$ electrons is to treat them as valence states using an
LSDA+U approach.\cite{anisimov} Tests performed
on the undistorted perovskite--like
cubic phase using  LSDA+U\cite{shick}  on the 4$f$ states showed
 that the resulting
 electronic structure is unchanged with respect to the ``4$f$-in-core"
 treatment.

\begin{figure}
\includegraphics[angle=90,scale=0.7,clip]{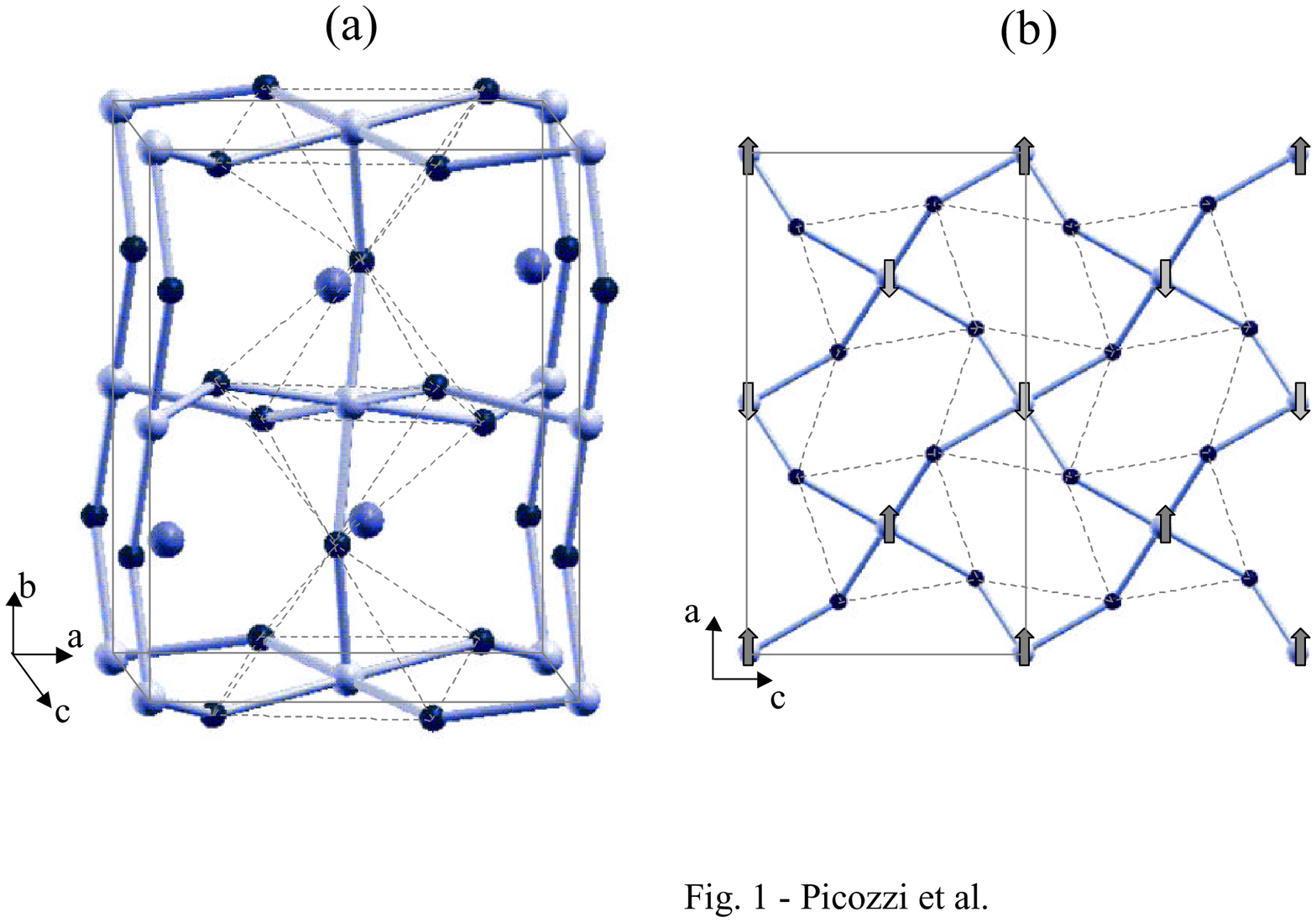}
\caption{(Color online) (a) The $Pnma$ orthorhombic cell of \rmo. Black, white and grey
spheres
represent O,  Mn and Tb (or Y or La or Ho) atoms, respectively.
(b) $ac$ in-plane view of MnO2 planes, showing the
large octahedral distortions and ``up-up-down-down" zig--zag
spin ordering. Solid lines
mark the in-plane projected unit--cell.}
\label{afmxcrys}
\end{figure}

The unit cell in the GdFeO$_3$-like orthorhombic phase shows the $Pnma$
symmetry
(20 atoms/unit-cell, choosing $b$ as the longest axis), with
enormous distortions with respect to the ideal cubic
perovskite (see Fig.\ref{afmxcrys}): 
due to the Jahn-Teller instability shown by
the Mn$^{3+}$ ion - with electronic configuration
$d^4$ ($t_{2g}^3\:e_g^1$) -, oxygen octahedrons
are highly distorted  and tilted (the average Mn-O-Mn angle is $\sim$ 145$^\circ$, $\sim$ 142$^\circ$
and $\sim$ 144$^\circ$ for \tmo, \hmo and \ymo, respectively, to be compared with the much
larger value of $\sim$ 155$^\circ$ in \lmo).
We recall that for \ymo (and for other R smaller
than Tb), the orthorhombic perovskite structure is no longer stable, the
hexagonal non-perovskite structure competing in stability. However, the
transition to the metastable orthorhombic $Pnma$ can be obtained by high
pressure synthesis\cite{alonso}.
The experimental lattice constants
and internal positions according to neutron diffraction data - very similar for the three compounds, \hmo
\ymo and \tmo - have been used
throughout the work\cite{alonso}. In order to separate chemical and structural
effects, we have also considered: {\em a}) \lmo, having less distorted
octahedrons and {\em b}) two ``artificial" systems, namely {\em i})
\ymo in the \lmo
structure and   {\em ii}) \lmo in the \ymo
structure (denoted in the following as [\ymo]$_{La}$ and [\lmo]$_{Y}$,
respectively).

\section{Results and discussion}

Let us first focus on \ymo, \tmo, \hmo and \lmo in their equilibrium structure.
In Table \ref{tabenergy}  we  report  the  total  energies  (calculated  per
formula-unit containing one Mn atom) with respect to  the  FM  phase. 
 If we restrict to the FM and AFM - A, C and G orderings -
({\em i.e.}
we neglect possible next-nearest-neighbors AFM coupling), 
the  situation  for \hmo, 
\tmo and \ymo is common to  the  well  studied  LaMnO$_3$,  where  the  most
stable magnetic configuration, as experimentally 
observed\cite{lamno3exp} and  theoretically
 confirmed\cite{ravindran,noriaki}, is the AFM-A type\cite{notaag}.  However, the inclusion
 of
 the AFM-E in the subset of considered
 magnetic configurations largely changes the
 scenario: interestingly,
among
the     collinear  magnetic   orderings,  {\em i}) \hmo show AFM-E as magnetic ground state, in excellent
agreement with experiments,  {\em ii})
both  \ymo  and  \tmo  show  the  AFM-E  type  as   stable
configuration and  {\em iii})  in
\lmo the AFM-E is also not too high in energy
 compared to the AFM-A stable configuration.  Indeed, this  confirms
 what was proposed via model--hamiltonians, {\em
i.e.} the occurrence of the AFM-E phase adjacent to the A-AFM in parameter space
and competing with the FM metallic phase as well\cite{dagotto}.  
The differences between \hmo, \tmo and \ymo - having very similar structural 
parameters - suggests
that, to some extent, the 4$f$ moment affects the Mn ordering. 
In order to further highlight  the  effect  of
the spin-polarization due to the 4$f$ Tb electrons,  we    considered  a
configuration in which these electrons are kept in the core,  but  they  are
equally shared between the majority and minority spins (4 in the  up  and  4
in the down channels, denoted as (4$\uparrow$,4$\downarrow$)), 
instead of the ``normal"  configuration  -  considered
so far - in which there are 7 and 1  electrons  in  the  up  and  down  spin
channels, respectively, denoted as (7$\uparrow$,1$\downarrow$)). 
The results show that the energy ordering of the
different phases in \tmo$^{(4\uparrow,4\downarrow)}$ is very similar to \ymo,
therefore suggesting
that the Mn--Mn coupling is appreciably affected by Tb 4$f$ electrons.
In particular, due to the geometric location of Tb
atoms ({\em i.e.} in between $ac$-planes), the main
differences induced by the treatment of Tb 4$f$ states occur in the
out-of-plane interactions, appreciably affected
by the 4$f$ 
``spin-polarized" or ``paramagnetic" electronic cloud in the
$(7\uparrow,1\downarrow)$ and $(4\uparrow,4\downarrow)$ case, respectively.

At this point, one might infer that the stability of the AFM-E is closely
related to
distortions (larger in \hmo, \ymo and \tmo compared to LaMnO$_3$). Indeed, this is
confirmed by our ``artificial" structures: in [\ymo]$_{La}$ the reduced
distortions - compared to \ymo in equilibrium - restore the stability of the
AFM-A (similar to \lmo), whereas the increased distortions in [\lmo]$_Y$
- compared to \lmo at equilibrium - make the AFM-E the ground state. Therefore, the stability of the AFM-E seems to be mainly connected with {\em structural} ({\em i.e.} octahedral distortions)
more than with {\em chemical} ({\em i.e.} identity of the rare-earth element) effects.

\begin{figure}
\includegraphics[angle=-90,scale=0.8,clip]{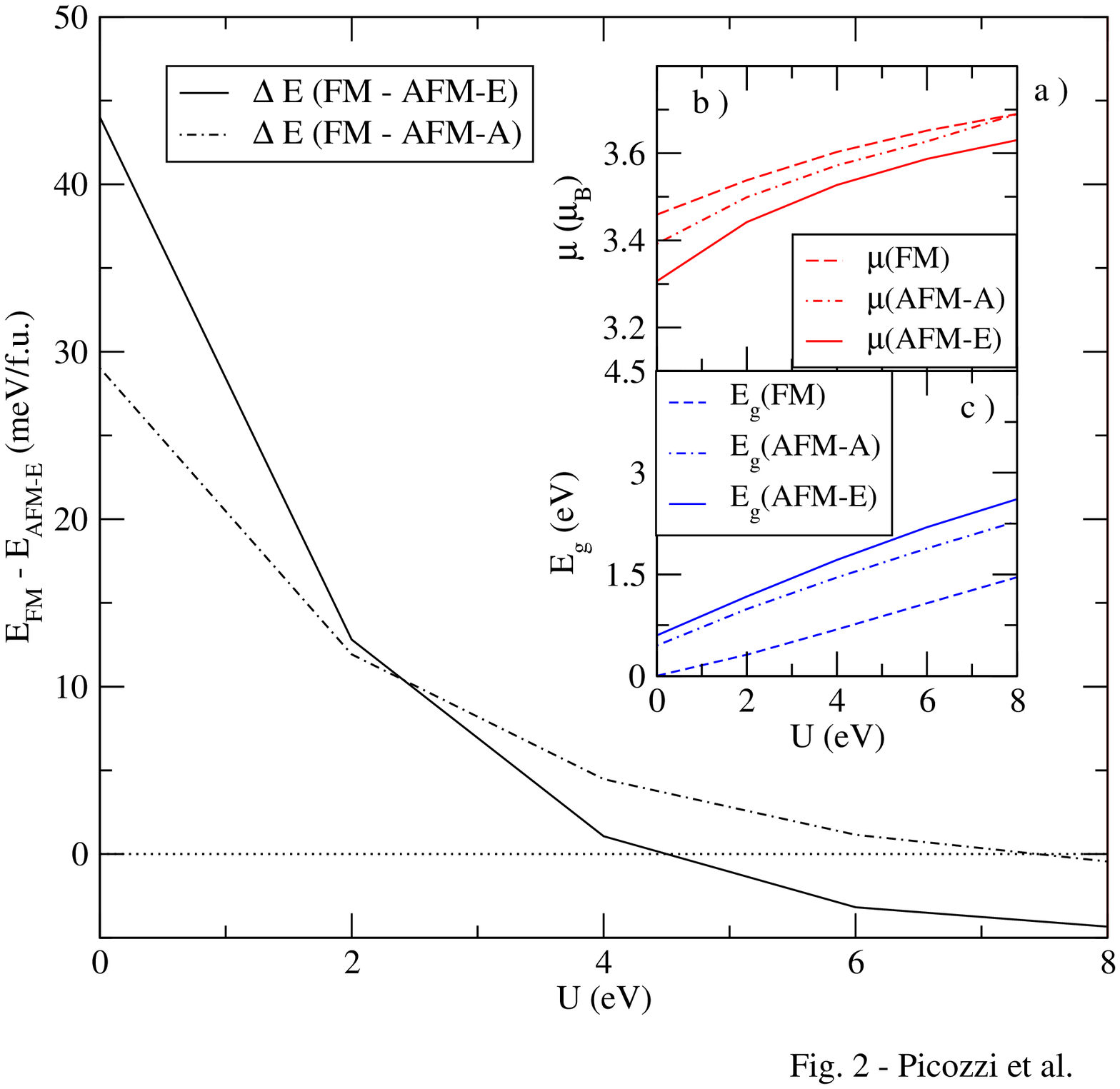}
\caption{(Color online) (a) Difference (in meV/formula-unit) between the FM and AFM-E (black solid line) and the FM and AFM-A (black dot-dashed line) as a function of the Coulomb parameter for \ymo. J is set to 0.15*U. The changes of (b) Mn magnetic moment and   (c) band-gap as a function of U are also shown for the different phases: FM (dashed line), AFM-A (dot-dashed line) and AFM-E (solid line).}
\label{ldaU}
\end{figure}

In order
to investigate the role of electronic correlation on the stability of the
different magnetic states, we performed some calculations for \ymo in the FM, AFM-A and AFM-E, according to an LSDA+U (or GGA+U) approach in the ``atomic-limit"\cite{anisimov} and varying the U value from 0 to 8 eV,
keeping the U/J ratio fixed and choosing J=0.15*U. In Fig.\ref{ldaU} a) we show the difference between total energies
in the AFM-A and FM
as well as in the AFM-E and FM. For a complete description of the electronic and magnetic structure, we also report the variation of the magnetic moment and of the band-gap as a function of the Coulomb parameter (see Fig. \ref{ldaU} b) and c), respectively). The bandgap and the magnetic moment both increase as a function of U, with a strong and rather weak dependence on the Coulomb parameter, respectively. This is expected, since LSDA+U tends to enhance the ``localization" of Mn $d$ states (thereby increasing the magnetic moment) and to push unoccupied states up in energy (thereby increasing the band-gap). As far as the magnetic ground state is concerned, our results show that the stability of the AFM-E and AFM-A is strongly affected by correlation effects: for values of U larger than 4 (7.5) eV, the FM is stabilized with respect to the AFM-E (AFM-A). 
We remark that the LSDA+U formalism might
overestimate the tendency towards ferromagnetism, as already
noted by Terakura and coworkers for \lmo \cite{solovyev} and consistently  with
our results in \ymo. In addition, to our knowledge and at present,  there are no spectroscopic works
for \ymo in the orthorhombic phase, on the basis of which one could extract the  value of the Coulomb parameter. So, the LSDA+U results are mainly shown to warn the reader that the magnetic ground state may be affected by correlation effects. We hope our work will stimulate further experimental works on distorted manganites to gain additional insights on the accuracy of bare GGA vs GGA+U in treating these systems. 

More information on the    magnetic  interactions  between
 Mn atoms can be gained by considering a Heisenberg 
Hamiltonian\cite{notaoo}  and  using
  this  model  to
fit our GGA-calculated total energies. In particular,  we  consider  the  first-
nearest-neighbor 
($J_{\parallel}^{nn}$) and second nearest-neighbor coupling 
($J_{\parallel}^{nnn}$) in the $ac$ plane, as well as  the  first-
and second-nearest-neighbour coupling out--of--plane 
($J_{\perp}^1$
and $J_{\perp}^2$).  
Before discussing quantitatively our results, let us first recall what happens in the well--understood \lmo, showing a smaller
\gdo-type distortion: the $nn$ FM superexchange (SE) between $e_g$ spins
competes with the AFM SE between $t_{2g}$ spins.\cite{noriaki}
 The delicate balance of these
two mechanisms crucially depends on the MnO$_6$ tilting and Jahn-Teller
distortions.\cite{alonso} Indeed, this is evidenced by our results: for the heavily distorted compounds, the FM $nn$ exchange parameters ($J_{\parallel}^{nn}\sim$-1-2 meV) are smaller in module
\cite{kimura1} than the much larger value in \lmo ($J_{\parallel}^{nn}\sim$-9 meV).
This can be explained considering that, upon large tilting ({\em
i.e.} in \hmo, \tmo and \ymo), the overlap between Mn 3$d$ and O 2$p$ orbitals is not
large enough to provide a strongly FM SE between Mn cations in the $ac$ plane.
In turn, this alters the delicate balance with
the $nnn$ SE
interactions through Mn-O-O-Mn paths along the $a$ direction.
Upon  significant \gdo-distortions, the
large $J_{\parallel}^{nnn}$  (of the order of few meV in all compounds) AFM coupling prevails, therefore stabilizing
 the AFM-E ordering. Furthermore, 
the out--of--plane coupling between $t_{2g}$ spins ($J_{\perp}^1$ of the order of few meV) keeps a strong AFM  character
along the R series, whereas  we obtain very small ($<$ 0.5 meV)
$J_{\perp}^2$ coupling  perpendicular
out--of--plane. 
Finally, the octahedral distortions also affect this
interplanar 
coupling ($J_{\perp}^1$ is smaller in \lmo than in distorted manganites), therefore weakening  the  assumption, very common in model 
hamiltonians, of
restricting the study to a 2D square lattice\cite{dagotto,kimura1}, 
neglecting out--of--plane effects. 

\begin{figure}[hbt]
\centering
\hspace{-1.25cm}
\includegraphics[scale=0.48,clip]{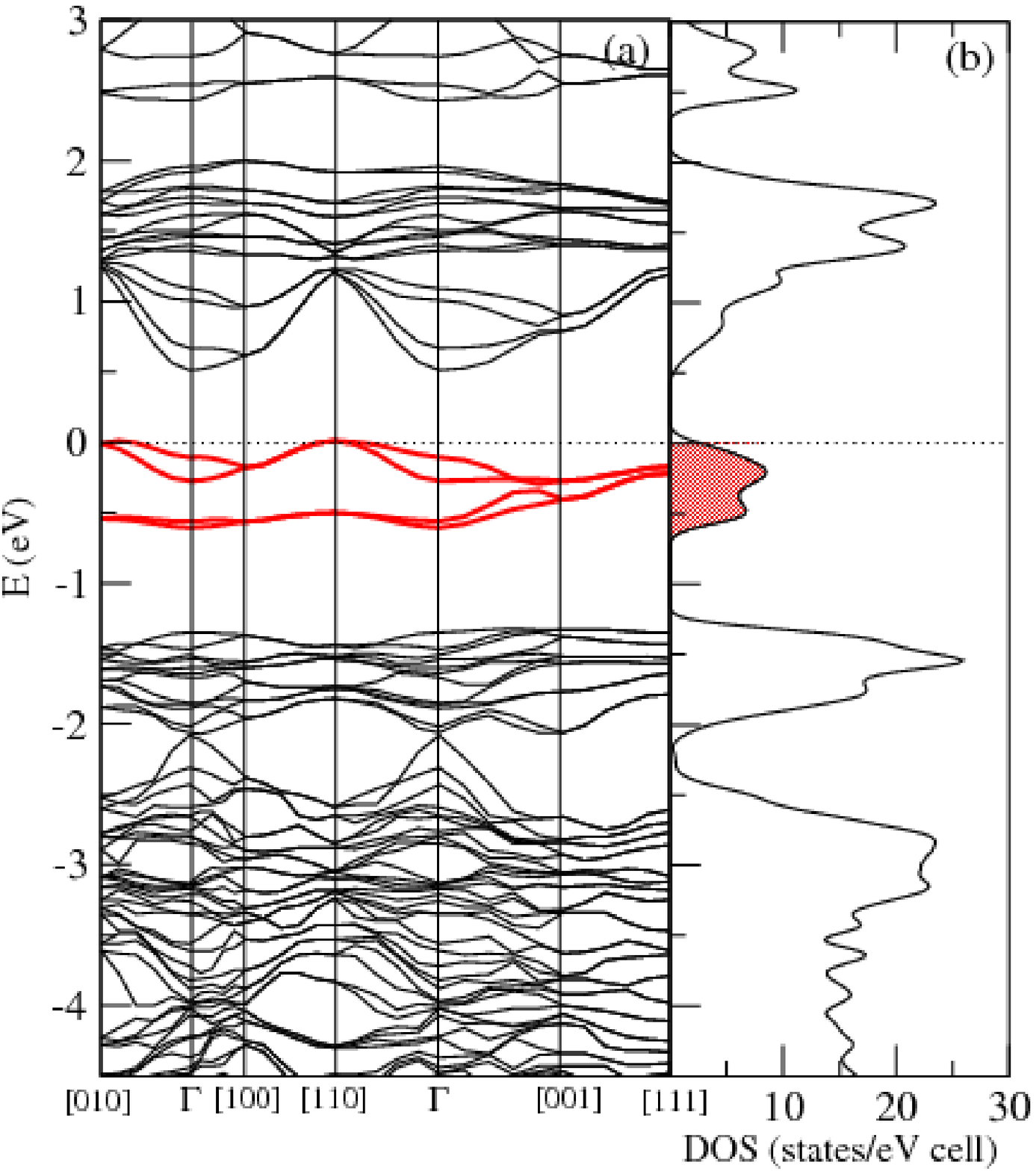}
\hfill
\hspace{-0.5cm}
\centering
\includegraphics[scale=0.43,clip]{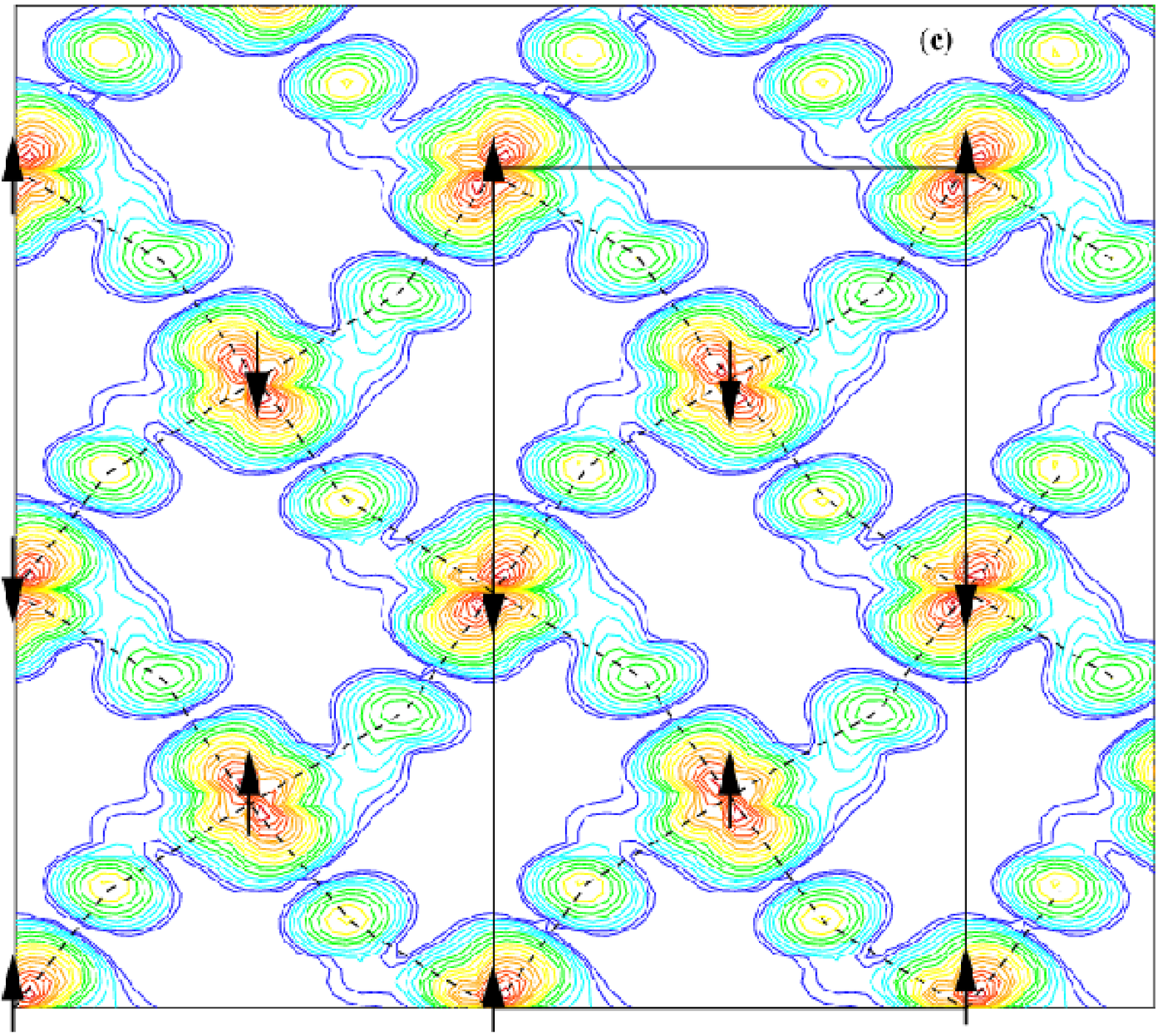}
\caption{(Color online) (a) Band-structure  and (b) total DOS for \ymo in the 
AFM-E magnetic configurations. (c) Total charge density contour plots
for occupied $e_g$ states.
Arrows denote the Mn spins; the solid lines mark the unit cell.}
\label{figorthodos}
\end{figure}

Let us now discuss the electronic and magnetic properties of the AFM-E
phase, in  terms
of magnetic moments,  band  structures  and  density  of  states.  
The GGA Mn local magnetic moment ($\sim$ 3.3 $\mu_B$)
is reduced with respect to  the
Hund's rule value (4$\mu_B$), reflecting the  hybridization  with 
O  atoms (slightly ferromagnetically polarized). 
The GGA-calculated total density of states and band-structure
for the AFM-E \ymo  are  reported
in Fig. \ref{figorthodos}.  Common to the well studied \lmo and shown by the
density of states projected on the muffin-tin spheres (not reported),
there  is  a  strong
hybridization between Mn $d$ and O $p$  states.  In  particular,  the  bands
lying just below $E_F$ (see highlighted states in Fig.\ref{figorthodos} (a) and
(b)) and separated by the rest  of  the  valence
band are  Mn-O  bands  (occupied  hybridized
$e_g$-like  states).  In agreement with the experimentally observed
insulating character\cite{kimura1},
the  resulting  band
structure shows an indirect band gap, with the conduction  band  minimum  at
$\Gamma$ and the valence band maximum at the Brillouin zone edge  along  the
[010] line and - more or less degenerately - along the [110]-$\Gamma$ line.
It is interesting to note that, even without  the
introduction of the on-site Coulomb U repulsion, 
the AFM-E configuration
 shows  an  energy  gap,  in
agreement with the experimental observation of an insulating behaviour.  When
comparing with the metallic character of the cubic perovskite-like systems (not
shown), despite their FM or AFM spin alignments, the
effect of the Jahn-Teller  distortion  and  of  the  octahedral  tilting  is
seen to result in a removal of states at the Fermi level by opening band  gaps.
\cite{ravindran} As shown in Fig.\ref{ldaU} c), the precise value of the band-gap is strongly affected by the inclusion
of correlation effects: for the extreme case of U = 8 eV, the gap increases up to $\sim$ 2.5 eV.
Finally, in order to have some hints on the orbital physics
in AFM-E, we plot in Fig.
\ref{figorthodos} (c) the \ymo total charge density of the occupied $e_g^{\uparrow}$
states (in the [-0.7;0] eV energy range with respect to $E_F$), which remarkably
shows the staggered ordering of the relevant $e_g$ orbital occurring in the Mn basal plane.

\section{Conclusions}

In summary, we have shown via density--functional--based methods
that in manganites with highly distorted MnO$_6$ octahedra, such as \hmo, \tmo and
\ymo, the insulating AFM-E
phase is stabilized. This is in excellent agreement with experiments, where the AFM-E  is observed for \hmo. The calculated exchange parameters show that \gdo-like
distortions - that increase along the rare-earth series - reduce the
in--plane nearest-neighbor FM
superexchange interaction along the Mn-O-Mn path. This interaction, along with
the delicate interplay with the AFM
next--nearest--neighbors superexchange along planar Mn-O-O-Mn paths
 is ultimately responsible for the
stabilization of the peculiar AFM-E spin arrangement.
Although the approximations made in this study (such as partially inaccurate
treatment of correlation effects and atomic relaxations, etc.)
should well be tested in the near
future, we hope that these results
 will stimulate more works - both from theory and
experiments - to ascertain the role played by the AFM-E magnetic configuration
in the manganites framework.

\begin{table*}

\caption{Calculated total energy difference (in meV/Mn) with respect to the
FM
phase. Recall that {\em i}) type-E consists of intraplane up-up-down-down and
interplane AFM coupling; {\em ii}) type-E$^*$ consists of intraplane up-up-down-down and
interplane FM coupling; {\em iii}) type-A consists of interplane AFM and
intraplane FM coupling; {\em iv}) type-C consists of intraplane AFM and
interplane FM coupling; {\em v}) type-G consists of both interplane  and
intraplane AFM coupling. Numbers in bold denote the ground state.}

\vspace{0.5cm}

\begin{tabular}{|c|c|c|c|c|c|c|c|}\hline \hline
 & \ymo & \hmo &\tmo$^{(7\uparrow,1\downarrow)}$ & \tmo$^{(4\uparrow,4\downarrow)}$ &
 \lmo  & [\ymo]$_{La}$ &  [\lmo]$_{Y}$ \\  \hline \hline
FM & 0 & 0& 0 & 0 & 0 & 0&0 \\\hline
AFM-E & {\bf -44} & {\bf -37} & {\bf -32} & {\bf -45} & -2  & +13& {\bf -61} \\ \hline
AFM-E$^*$ & -23 & -12 &-11 & -30 & +2 & +15& -34 \\ \hline
AFM-A & -29 & -16 & -8  & -21 & {\bf -17} & {\bf -7} & -38\\\hline
AFM-C & -10 & -2 & +17  & +9 & +58 & +78& -33\\\hline
AFM-G & -24 & -14 &+4 & -17 & +64 & +88& -52\\ \hline \hline

\end{tabular}

\label{tabenergy}

\end{table*}




 



\begin{thebibliography}{100}
\bibitem{notaag} Incidentally, we note that our results for \lmo are in overall
good agreement with previous first--principles calculations, such as
Refs.\protect{\cite{ravindran,solovyev}}.
\bibitem{notaoo} 
 Model-hamiltonians for manganites
 generally include an orbital-ordering term in addition to spin-spin coupling,
   which, however,
goes beyond the scope of the present work.
\end{thebibliography}
\end{document}